# Parallelization of Maximum Entropy POS Tagging for Bahasa Indonesia with MapReduce

Arif Nurwidyantoro[1] and Edi Winarko[2]

[1] Computer Science and Electronics Department, Universitas Gadjah Mada
Yogyakarta, 55281, Indonesia
*arifn@mail.ugm.ac.id*

[2] Computer Science and Electronics Department, Universitas Gadjah Mada
Yogyakarta, 55281, Indonesia
*ewinarko@ugm.ac.id*

**Abstract**
In this paper, MapReduce programming model is used to parallelize training and tagging proceess in maximum entropy part of speech tagging for Bahasa Indonesia. In training process, MapReduce model is implemented dictionary, tagtoken, and feature creation. In tagging process, MapReduce is implemented to tag lines of document in parallel. The training experiments showed that total training time using MapReduce is faster, but its result reading time inside the process slow down the total training time. The tagging experiments using different number of map and reduce process showed that MapReduce implementation could speedup the tagging process. The fastest tagging result is showed by tagging process using 1,000,000 word corpus and 30 map process.

## 1. Introduction

Part of speech (POS) tagging is the task of labeling (or tagging) each word in a sentence with its appropriate part of speech [1]. POS tagging is considered as one of preliminary task on natural language processing. POS tagging itself is an essential tool to various natural language processing applications, such as word disambiguation, parsing, question answering, and machine translation.

In natural language processing researches, data size matters. Researches showed that more data led to better accuracy [2]. This led to the suggestion to increase the training data and reduces the focus of research on the comparison of training methods using small-sized data [3].

MapReduce is a programming model and an associated implementation for processing and generating large data sets [4]. MapReduce has the facilities to handle constraints in parallel processing such as hardware failure and data usage from multiple sources. MapReduce library already has features to process text documents and support cloud computing platform.

The use of MapReduce for POS tagging has been done in English using Infinite HMM [5]. MapReduce has never been used for POS tagging using Maximum Entropy approach for Bahasa Indonesia. The utilization of MapReduce in POS tagging is expected to enhance scalability in large data processing.

## 2. Related Works

The research about Maximum Entropy in POS tagging conducted for the first time by Ratnaparkhi [6]. He created statistical model from training process using anotated corpus. This model uses contextual features to predict POS anotation in unanotated corpus. Tautanova and Manning [7] then added information sources in Maximum Entropy model to increase the accuracy of POS tagging to unknown words. The added features are word capitalization, features for the disambiguation of the tense forms of verbs, and features disambiguating particles from prepotitions and adverbs. Van Gael et al. [5] using MapReduce to optimize the computation process on their POS tagging research. They used Infinite HMM POS tagging for English language.

Researches about POS tagging for Bahasa Indonesia already conducted using various approaches such as Brill's transformational rule based [8], Maximum Entropy and Conditional Random Fields [9], and Hidden Markov Model [10]. The highest accuracy in Maximum Entropy POS tagging experiments is 97.17% [9].

## 3. Background Theory

### 3.1 Maximum Entropy POS Tagging

Maximum Entropy method assigns probability value for each anotation according to contextual information in training corpus [7,9]. The probability model is defined as [6] shown in Eq. (1) with $h$ as "histories" or word and its context, $t$ is tag from set of possible tags, $\pi$ is a normalization constant, $\{\mu, \alpha_1, ..., \alpha_k\}$ are the positive model parameters, and $\{f_1, ..., f_k\}$ are known as "features", where $f_j(h,t) \in \{0,1\}$. Each parameter $\alpha_j$ corresponds to a feature $f_j$.

$$p(h,t) = \pi\mu \prod_{j=1}^{k} \alpha_j^{f_j(h,t)} \quad (1)$$

The model parameters must be set so as to maximize the entropy of probability distribution subjects to the constraints imposed by the value of $f_j$ feature functions observed from training data [9]. These parameters usually trained using Generalized Iterative Scaling (GIS) algorithm [6]. However, Improved Iterative Scaling (IIS) algorithm also can be used to improve the slow convergence of GIS [11].

Tagging process is done by using probability of a tag sequence $t_1 ... t_n$ given a sentence $w_1 ... w_n$ [6,7] as shown in Eq. (2).

$$p(t_1 ... t_n | w_1 ... w_n) \approx \prod_{i=1}^{n} p(t_i | h_i) \quad (2)$$

### 3.2 MapReduce

MapReduce is a programming model and an associ-ated implementation for processing and generating large data sets [4]. MapReduce uses functional programming model consists of map and reduce functions. Both of these functions are defined by user and processed in parallel.

Map function takes an input of key/value pair and produces a set of intermediate key/value pairs. The MapReduce library then groups together all intermediate values associated with the same intermediate key and passes them to the reduce function [4].

Reduce function receives an intermediate key and all the values associated with it. The function merges together these values to form possibly a smaller set of values [4]. Altough MapReduce uses simple functions, there are many data processing tasks that could be expressed using this model [4] as showed in Table 1.

Table 1: MapReduce examples

| Case | Map Output | Reduce Output |
|---|---|---|
| distributed grep | <pattern, line of documents> | <pattern, line of documents> |
| count of URL acces frequency | <URL, 1> | <URL, total count> |
| reverse web link graph | <target URL, page source> | <target URL, list(page source)> |
| term-vector per host | <hostname, term vector> | <hostname, term vector> |
| inverted index | <word, document ID> | <word, list(document ID)> |
| distributed sort | <key, record> | <key, record> |

## 4. Parallelization of Maximum Entropy POS Tagging

Automatic POS tagging usually involves training and tagging process. The training process takes manually anotated corpus as input to find a model that can be used to automatically labeling unanotated corpus. Meanwhile, the tagging process uses model, created from training process, to labels appropriate part of speech to each word in unanotated corpus

The parallelization techniques in training and tagging process are done by modifying Stanford POS tagger library. The parallelization system architecture is showed in Figure 1. The training process consists of several processes, namely the process of forming dictionary, tagtoken, histories, and features, and also the IIS algorithm process. These processes and its MapReduce parallelization techniques are described as follows.

Dictionary is generated by creating a list of words and its associated tags and its tag's frequency from training corpus. Map function is used to separate word and its tag from anotated corpus, produces word and its tag as intermediate key/value pairs. Reduce function collects tags associated with the same word and count each tag's frequency.

Tagtoken is generated by creating a list of tag and all the words associated with it from training corpus. Map function is used to separate word and its tag, produces tag and its associated word as intermediate key/value pairs. Reduce function collects words associated with the same tag.

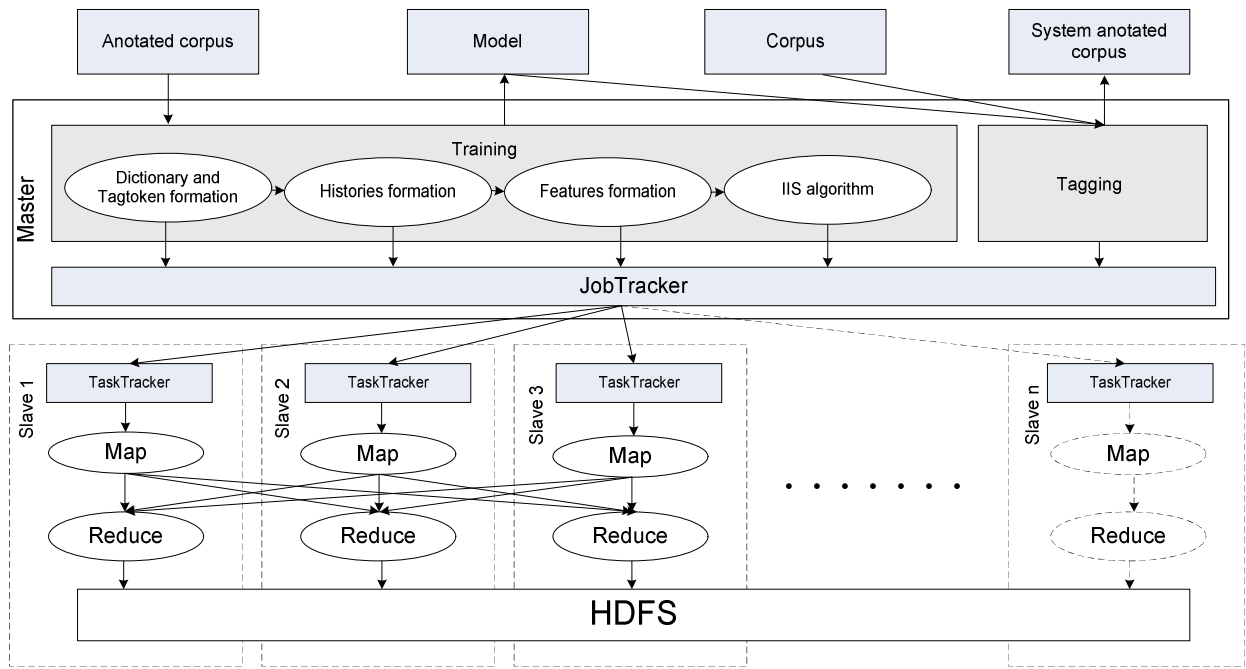

Fig. 1 System architecture

Histories records word and tag position along with all word-tag pairs in a sentence. Parallelization is done by partitioning training corpus and forms histories in different nodes. Map function separate word and its anotation, then list word-tag pairs in a sentence. Word position and word-tag pairs list is used as histories. Reduce functions just emits the map outputs unchanged.

Features are generated from histories information based on predefined features template. Map function is used to generate features according to features template. Reduce function collects features from map outputs.

The IIS algorithm is used to count the weight parameter of each features. Map function is used to count weight parameter changes for every features in parallel. Reduce function just emits the map outputs. MapReduce process is done iteratively according to iteration in IIS.

Parallelization in tagging process is done by partitions the unanotated corpus and gives label to each partition in different nodes. Map function gives part of speech label to words in documents. Reduce function collects and sorts anotated sentences from map function outputs.

## 5. Experiments

The experiments is conducted in training and tagging process. These experiments were aimed to compare both processes with and without MapReduce parallelization. The experiments used different parameters, such as corpus size, number of nodes, and also number of map and reduce process. The experiments used Hadoop MapReduce library.

### 5.1 Training Experiments

Table 2: Training experiment parameters

| Parameter | Value |
| --- | --- |
| arch | generic |
| learnClosedClassTags | true |
| closedClassTagsThreshold | 10 |
| curWordMinFeatureThresh | 2 |
| tagSeparator | / |
| search | iis |
| iterations | 500 |

Anotated corpus used in training experiments are 12,000 words corpus from Wicaksono and Purwarianti research [10] and also 100,000 words and 1,000,000 words corpus from PanLocalization. The experiments parameters is based on Stanford POS tagger parameters as showed in Table 2. The experiment uses generic architecture that not depend on language. Closed class tags is automatically learned if a tag has frequency less than 10 times in the corpus. Features are generated only for words that appears more than two times in the training corpus using feature templates showed in Table 3   The algorithm used to determine the weight parameters for each features is IIS with 500 iterations.

Table 3: Feature templates

| No. | Features | |
|---|---|---|
| 1. | $w_i$ | & $t_i$ |
| 2. | $w_{i-1}$ | & $t_i$ |
| 3. | $w_{i+1}$ | & $t_i$ |
| 4. | $t_{i-1}$ | & $t_i$ |
| 5. | $t_{i-1} t_{i-2}$ | & $t_i$ |
| 6. | $t_{i-1} w_i$ | & $t_i$ |
| 7. | $w_{i-1} w_i$ | & $t_i$ |

The MapReduce training experiments conducted using three nodes with 30 maps and 6 reduces. The training experiments showed that dictionary, tagtoken, histories, and features generation using MapReduce gave the same results with experiments without using MapReduce. The IIS algorithm experiments using MapReduce showed different results from the one without using MapReduce. This happened because there are differences in features updates sequence. The parameter update for one feature affect the other features. This make IIS algorithm could not be parallelized using MapReduce. The differences made MapReduce modification in IIS algorithm not used in training performance experiments.

Table 4: MapReduce training process time

| Job MapReduce | 12,000 words | 100,000 words | 1,000,000 words |
|---|---|---|---|
| Dictionary | 34 seconds | 35 seconds | 53 seconds |
| Tagtoken | 36 seconds | 37 seconds | 43 seconds |
| Histories | 43 seconds | 55 seconds | 133 seconds |
| Features | 53 seconds | 77 seconds | 337 seconds |

Table 4 shows MapReduce time for dictionary, tagtoken, histories, and features creation. The table shows that the dictionary and tagtoken creation time did not increase significantly with the increase of corpus size. The histories and features creation time increase significant along with the corpus size. This happened because bigger corpus create more histories and features.

Table 5 shows time differences between training using Stanford POS tagger with and without MapReduce modifications. The result shows that MapReduce training time using 12.000 words corpus is slower than training time without modifications. But using larger corpus MapReduce training time is faster than without modifications. Training time in Table 5 did not count the reading time from distributed file system. For every MapReduce process, the system should read MapReduce results from file system. This is the characteristic of Hadoop MapReduce library.

Table 5: Training time

| System | 12,000 words | 100,000 words | 1,000,000 words |
|---|---|---|---|
| Stanford POS tagger | 358.1 seconds | 3,290.2 seconds | 58,219.7 seconds |
| Stanford POS tagger + MapReduce | 389 seconds | 3,279.3 seconds | 58,000.1 seconds |

Table 6: Total training time

| System | 12,000 words | 100,000 words | 1,000,000 words |
|---|---|---|---|
| Stanford POS tagger | 358.1 seconds | 3,290.2 seconds | 58,219.7 seconds |
| Stanford POS tagger + MapReduce | 503.2 seconds | 3,596.7 seconds | 59,492.5 seconds |

The total training time involving reading results from file system is showed in Table 6. The MapReduce training time is slower than without modifications. Apparently, the reading time is larger than the time difference between training with MapReduce and without MapReduce. To test the accuracy of the system, we use manually anotated corpus consists of 6,348 words. The accuracy results is shown in Table 7. The results shows that MapReduce modifications in training process did not change the model accuracy.

Table 7: Accuracy results

| System | 12,000 words | 100,000 words | 1,000,000 words |
|---|---|---|---|
| Stanford POS tagger | 73.70 % | 66.48 % | 68.49 % |
| Stanford POS tagger + MapReduce | 73.70 % | 66.48 % | 68.49 % |

### 5.2 Tagging Experiments

The text documents used in tagging experiments were obtained from various news sites consists of 10,000 to 1,000,000 words. There are three experiments for tagging process, each using different nodes and parameters. The models used for tagging experiments are the models from training experiments, created from training corpus consists of 12,000, 100,000, and 1,000,000 words.

The first tagging experiment is conducted using one node without MapReduce. The results in Figure 2 shows that tagging using 12,000 words model is much slower than using models from bigger corpus.

Figure 4 shows that using model from bigger corpus also gave the best time performance.

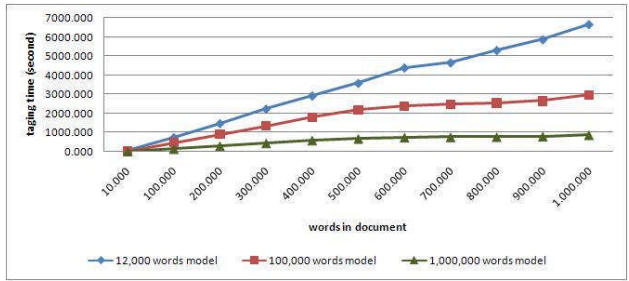

Fig. 2 Tagging time using one node.

The second tagging experiment is conducted using three nodes three maps and one reduce process. In this experiment, the tagging experiment documents are splitted into three parts. Each parts are tagged parallely and combined using one reduce process. The results in Figure 3 also shows that the tagging time using model from bigger corpus is faster than using model from smaller corpus.

The third tagging experiment is conducted in three nodes using 30 maps and 6 reduces processes. In this experiments, the tagging text documents are splitted into 30 parts and processed parallely in three nodes. The results then combined using 6 reduces processes. The results in

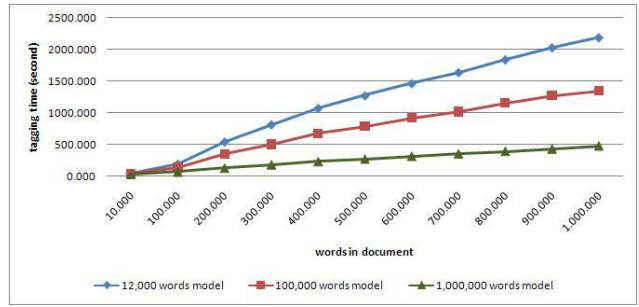

Fig. 3 Tagging time using 3 node, 3 maps, 1 reduce.

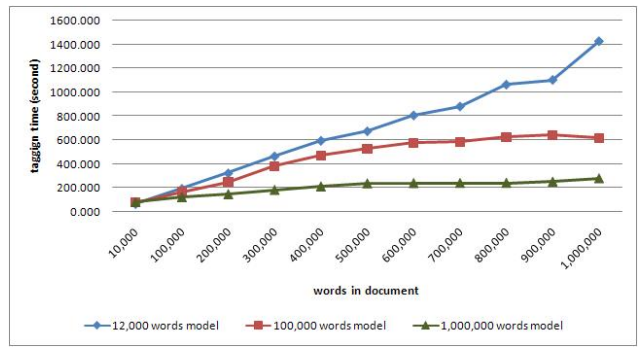

Fig. 4 Tagging time using 3 node, 30 maps, 6 reduce.

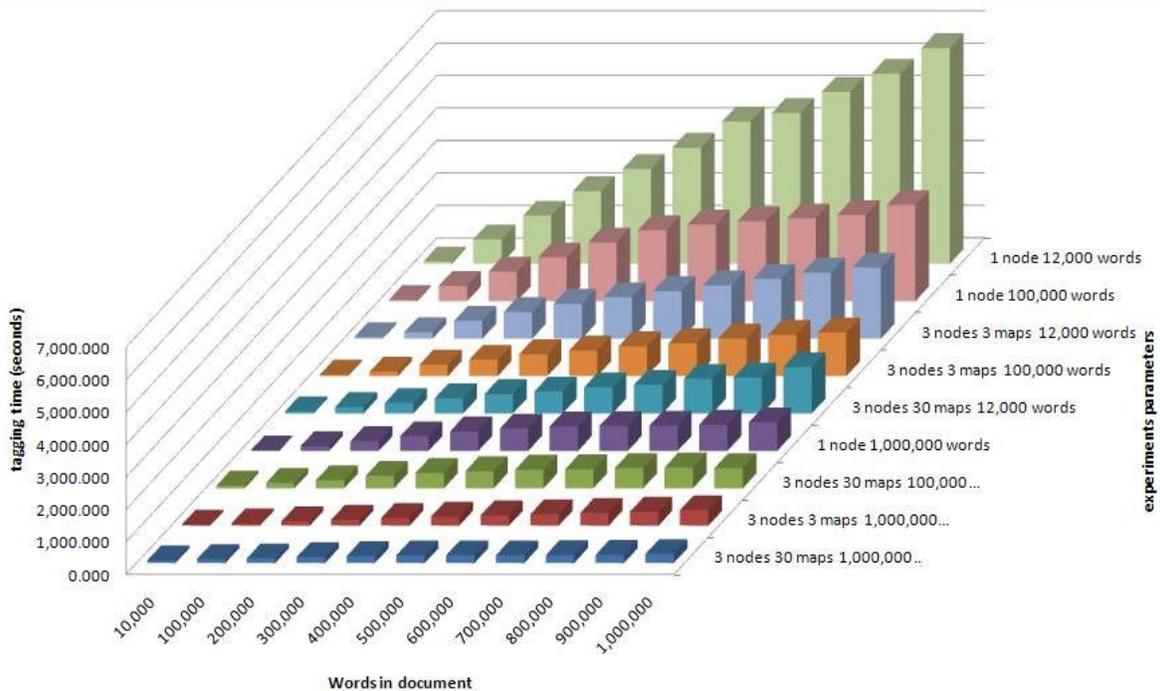

Fig. 5 Tagging time comparisons

The results of all tagging experiments were compared to see which parameter gives the best performance. Figure 5 showed the tagging time comparisons. The figure shows tagging using three nodes thirty maps and six reduces using model from 1,000,000 words gives the fastest tagging process. The slowest tagging process is the one which used single node and 12,000 words training corpus. In general, with the increasing text sizes, tagging time using three nodes did not increase significantly compared to tagging time using single node.

Figure 5 showed that tagging experiments using larger training corpus sizes give the best time. This happened because tagging process requires access to dictionary references. When a word cannot be found in the dictionary references, tagging process must consider all the tags in the tagset. Larger training corpus size has larger words in dictionary, so that there are many words found in it. This makes tagging using model from bigger training corpus faster than tagging using model from smaller training corpus.

## 6. Conclusions and Future Works

The experiments showed that MapReduce modifications in training process Maximum Entropy POS tagger is essentially hastened the process. However, reading result's time by Hadoop MapReduce library made total training time slower than training process without MapReduce. In tagging process, MapReduce implementation took less time than without MapReduce. The tagging experiments also showed that tagging using model from bigger corpus is faster than model from smaller corpus. The parallelization using MapReduce could enhance the performance of Maximum Entropy POS tagging.

In the next research, we will create model from bigger corpus to find out the influences of corpus size towards time and accuracy of POS tagger. Research can also be done using more nodes of computer such as in cloud computing platforms, like Amazon EC2. We also plan to use different parameter estimation algorithms, such as Quasi-Newton or Conjugate Gradient in place of IIS.

**Arif Nurwidyantoro** received his bachelor degree from Institut Pertanian Bogor, Indonesia, and master degree from Universitas Gadjah Mada, Indonesia, both in Computer Sciences. He currently works as teaching assistants at Universitas Gadjah Mada. He has interest in data mining, especially text and web mining, and also in large data processing.

**Edi Winarko** received his bachelor degree in Statistics from Universitas Gadjah Mada, Indonesia, M.Sc in Computer Sciences from Queen University, Canada, and Ph.D in Computer Sciences from Flinders University, Australia. He currently works as lecturer at Department of Computer Sciences and Electronics, Faculty of Mathematics and Natural Sciences, Universitas Gadjah Mada. His research interests are data warehousing, data mining, and information retrieval. He is a member of ACM and IEEE.